\documentclass[aps,prl,twocolumn,preprintnumbers,superscriptaddress,amsmath,amssymb]{revtex4-1}

\usepackage[colorlinks,linkcolor=blue]{hyperref}

\usepackage[normalem]{ulem}
\usepackage{graphicx}
\usepackage{subfigure}
\usepackage{epsfig}
\usepackage{xcolor}
\usepackage{dcolumn}
\usepackage{bm}
\usepackage{ulem}
\usepackage{color}
\usepackage{amsmath}
\usepackage{amsfonts}
\usepackage{amssymb}

\usepackage{epstopdf}

\begin{document}

\title{Experimental evidence of type-II Dirac fermions in PtSe$_2$}

\author{Kenan Zhang}
\affiliation{State Key Laboratory of Low Dimensional Quantum Physics and Department of Physics, Tsinghua University, Beijing 100084, China}

\author{Mingzhe Yan}
\affiliation{State Key Laboratory of Low Dimensional Quantum Physics and Department of Physics, Tsinghua University, Beijing 100084, China}

\author{Haoxiong Zhang}
\affiliation{State Key Laboratory of Low Dimensional Quantum Physics and Department of Physics, Tsinghua University, Beijing 100084, China}

\author{Huaqing Huang}
\affiliation{State Key Laboratory of Low Dimensional Quantum Physics and Department of Physics, Tsinghua University, Beijing 100084, China}

\author{Masashi Arita}
\affiliation{Hiroshima Synchrotron Radiation Center, Hiroshima University, Higashihiroshima, Hiroshima 739-0046, Japan}

\author{Zhe Sun}
\affiliation{National Synchrotron Radiation Laboratory, University of Science and Technology of China, Hefei, Anhui 230029, China}

\author{Wenhui Duan}
\affiliation{State Key Laboratory of Low Dimensional Quantum Physics and Department of Physics, Tsinghua University, Beijing 100084, China}
\affiliation{Collaborative Innovation Center of Quantum Matter, Beijing, China}

\author{Yang Wu}
\affiliation{Department of Physics and Tsinghua-Foxconn Nanotechnology Research Center, Tsinghua University, Beijing, 100084, China}

\author{Shuyun Zhou}
\altaffiliation{Correspondence should be sent to syzhou@mail.tsinghua.edu.cn}
\affiliation{State Key Laboratory of Low Dimensional Quantum Physics and Department of Physics, Tsinghua University, Beijing 100084, China}
\affiliation{Collaborative Innovation Center of Quantum Matter, Beijing, China}

\date{\today}

\begin{abstract}
Topological semimetals have attracted extensive research interests for realizing condensed matter physics counterparts of three-dimensional Dirac and Weyl fermions, which were originally introduced in high energy physics. Recently it has been proposed that type-II Dirac semimetal can host a new type of Dirac fermions which break Lorentz invariance and therefore does not have counterpart in high energy physics. Here we report the electronic structure of high quality PtSe$_2$ crystals to provide direct evidence for the existence of three-dimensional type-II Dirac fermions. A comparison of the crystal, vibrational and electronic structure to a sister compound PtTe$_2$ is also discussed. Our work provides an important platform for exploring the novel quantum phenomena in the PtSe$_2$ class of type-II Dirac semimetals.

\end{abstract}

\maketitle

\begin{figure*}
  \centering
  \includegraphics[width=12cm]{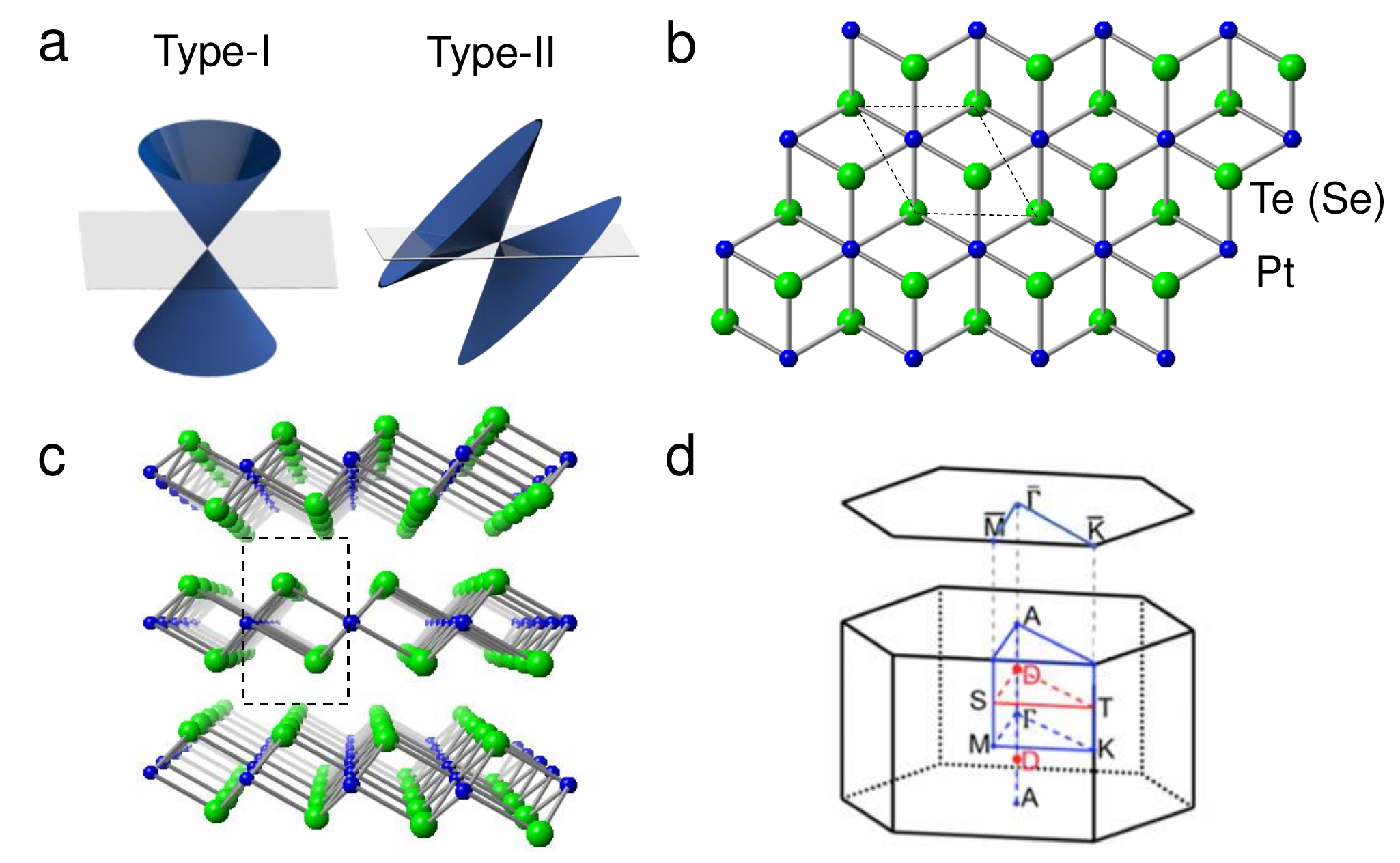}
  \caption{\textbf{: Crystal structure and Brillouin zone of PtSe$_2$ ( PtTe$_2$ ).} (a) Schematic drawing of type-I and type-II Dirac fermions. (b,c) Top and side views of the crystal structure of PtSe$_2$ (PtTe$_2$). Green balls are Se (Te) atoms, and blue balls are Pt atoms. The unit cell is indicated by black dashed lines. (d) BZ of bulk and the projected surface BZs of a (001) surface. The blue dots indicate the high-symmetry points of the BZ, and the red dots highlight the 3D Dirac point position.}\label{Fig1}
\end{figure*}

\begin{figure*}
  \centering
  \includegraphics[width=16.8cm]{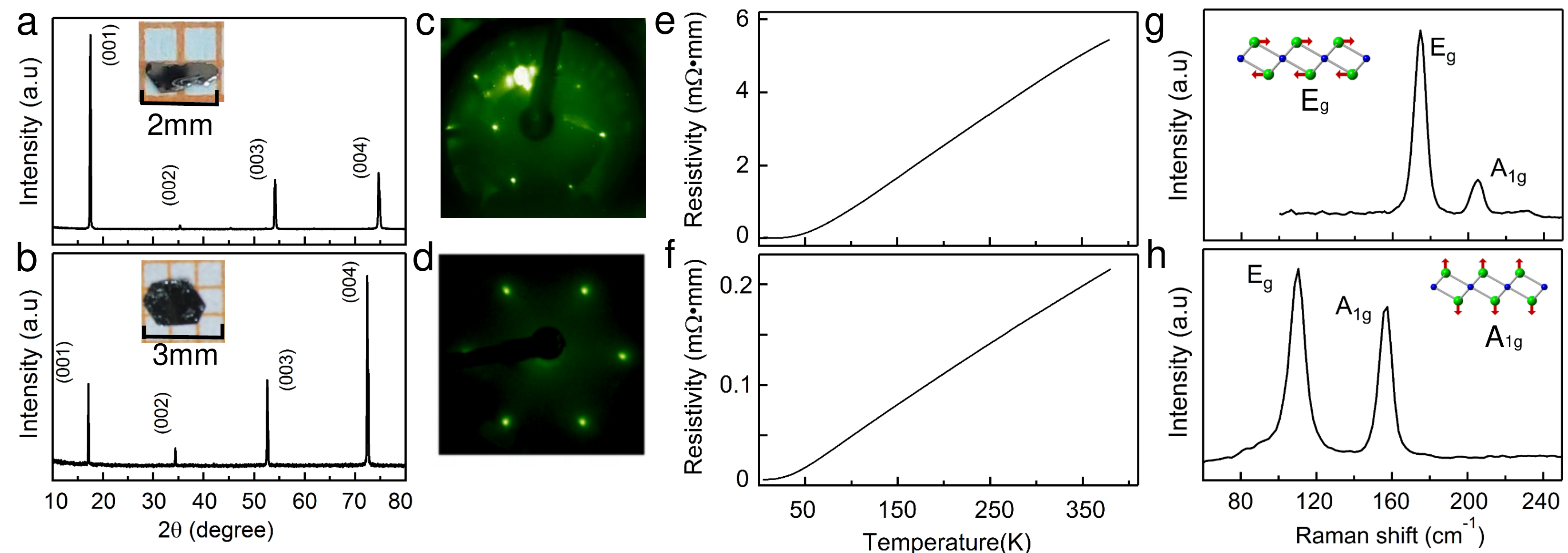}
  \caption{\textbf{: Characterization of PtSe$_2$ as compared to PtTe$_2$.} (a,b) X-ray diffraction patterns of PtSe$_2$ and PtTe$_2$ measured at room temperature. The inset shows a picture of the few-millimetre-size single crystal. (c,d) LEED pattern of PtSe$_2$ and PtTe$_2$. (e,f)Temperature dependent resistivity of PtSe$_2$ and PtTe$_2$. (g,h) Raman spectrum of PtSe$_2$ and PtTe$_2$ measured at room temperature. The inset illustrates vibrational modes in PtSe$_2$ (PtTe$_2$) with the same color code used in Fig. \ref{Fig1}. Arrows guide the vibration directions.}\label{Fig2}
\end{figure*}

\begin{figure*}
  \centering
  \includegraphics[width=16cm]{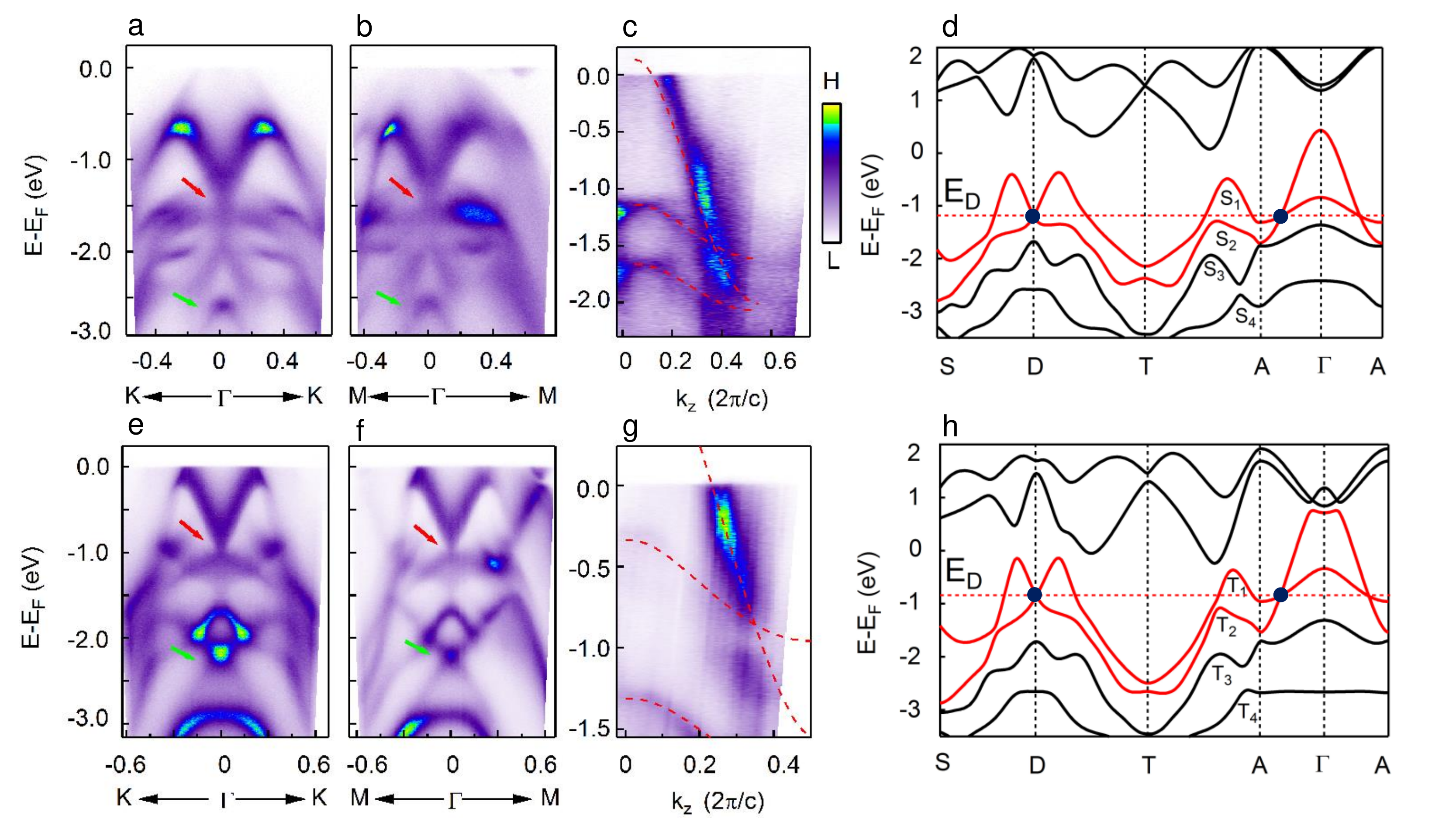}
  \caption{\textbf{Band structure of type-II Dirac cone in PtSe$_2$ and PtTe$_2$.} (a,b) PtSe$_2$ band dispersions along the $\Gamma$-K and $\Gamma$-M, respectively at a photon energy of 23.5 eV (k$_z$=0.37 c$^*$). (c) k$_z$ dispersions of PtSe$_2$ measured at k$_\parallel$ = 0.  Red broken lines are calculated dispersions for comparison. (e,f) PtTe$_2$ band dispersions along the $\Gamma$-K and $\Gamma$-M, respectively at a photon energy of 22 eV (k$_z$=0.35 c$^*$). (g) k$_z$ dispersions of PtTe$_2$ measured at k$_\parallel$ = 0. (d,h) Calculated band dispersion of PtSe$_2$ and PtTe$_2$ along the in-plane direction S-D-T and out-of-plane direction A-D-$\Gamma$-D-A through the Dirac point. }\label{Fig3}
\end{figure*}

\maketitle

\section{I. INTRODUCTION}
 Dirac and Weyl fermions were originally introduced in high energy physics \cite{Weyl1929pnas}. Their  counterparts in condensed matter physics have been realized in three-dimensional (3D) Dirac and Weyl semimetals. Such topological semimetals exhibit a rich variety of novel phenomena, such as negative magnetoresistance (MR) \cite{OngNa3Bi,OngCd3As2,ChenTaAsprx}, chiral magnetic effects \cite{Zyuzin2012Topological}, and quantum anomalous Hall effect \cite{Liu2015Erratum}. Dirac and Weyl semimetals can be classified into type-I and type-II, depending on whether the Lorentz invariance is preserved or not. For type-I Dirac \cite{ChenYLScience,HasanCd3As2} and Weyl semimetals \cite{WengHMprx,DingHTaAsprx,HasanTaAssci} which obey Lorentz invariance, massless Dirac fermions with linear dispersions are expected at the Dirac or Weyl points.
Type-II Dirac and Weyl fermions  \cite{BernevigNature} emerge at the topologically protected touching points of electron and hole pockets, and they show highly tilted Dirac cones along certain momentum direction (see schematics in Fig.~\ref{Fig1}(a)), thereby breaking the Lorentz invariance \cite{Deng2016Experimental,HasanMTnc,KaminskiMTnm}. The anisotropic electronic structure can also lead to anisotropic MR, and negative MR is expected only along directions where the cones are not tilted enough to break the Lorentz invariance \cite{BernevigNature,Zyuzinarxiv,berryphasePdTe2}. Type-II Dirac semimetal can be tuned to a Weyl semimetal or topological crystalline insulator when the crystal symmetry or time reversal symmetry is broken \cite{Burkov2011Weyl,Hal2012Time}, and therefore they are ideal candidates for investigating topological phase transitions and potential device applications.

 Recently, PtTe$_2$ has been reported to be a Lorentz violating type-II Dirac semimetal \cite{Yan2016Lorentz}. Its sister compound PtSe$_2$ has also been predicted to be a type-II Dirac semimetal  \cite{Huang2016Type}. Although monolayer PtSe$_2$ film has been shown to exhibit interesting electronic properties \cite{PtSe2N} and helical spin texture with spin-layer locking \cite{PtSe2nc}, so far research in bulk PtSe$_2$ crystals has been limited due to the lack of high quality single crystals. Here we report the growth and characterization of high quality PtSe$_2$ single crystal, and provide direct experimental evidence for type-II Dirac fermions from angle-resolved photoemission spectroscopy (ARPES). A comparison of PtSe$_2$ with PtTe$_2$ in crystal, vibrational and electronic structure is also discussed.

\section{II. METHODS}
Although high quality PtTe$_2$ single crystals can be directly synthesized by a self-flux method as reported previously \cite{Yan2016Lorentz}, PtSe$_2$ single crystals cannot be grown using the same method. Instead, we need to use chemical vapor transport (CVT). Precursor of polycrystalline PtSe$_2$ was synthesized by directly heating the stoichiometric mixture of high-purity Pt granules (99.9\%, Alfa Aesar) and Se ingot (99.99\%, Alfa Aesar) at 1,173 K in a vacuum-sealed silica ampoule for 3 days. The polycrystalline PtSe$_2$ was then recrystallized by the CVT method using SeBr$_4$ (99\%, Aladdin, with a concentration of $\leq $2.7 mg/ml) as the transporting agent. The growth rate is much slower than PtTe$_2$. After 3 weeks, PtSe$_2$ single crystals of a few millimeters in size with shiny surfaces can be obtained.

The band structures were carried out using density functional theory (DFT) as implemented in the Vienna ab initio simulation package \cite{VASPcms}. We employ the Perdew-Burke-Ernzerhof–type generalized gradient approximation \cite{PBEprl} and the projector augmented wave (PAW) method \cite{Bl1994Improved}. A plane wave basis set with a default energy cutoff and an $18\times 18\times 11$ k-point mesh were used. Spin-orbit coupling (SOC) is included self-consistently within the DFT calculation.

ARPES measurements were taken at BL13U of Hefei National Synchrotron Radiation Laboratory, BL9A of Hiroshima Synchrotron Radiation Center under the proposal No.15-A-26 and our home laboratory. The crystals were cleaved \textit{in-situ} and measured at a temperature of T$\approx$20 K in vacuum with a base pressure better than 1$\times$10$^{-10}$ Torr.

\section{III. RESULTS}
As its isostructural counterpart PtTe$_2$, PtSe$_2$ crystallizes in the centrosymmetric CdI$_2$-type structure with space group P$\overline{3}$m1 (No.~164) and point group D$_{3d}$ (-3m). The structure can be regarded as the hexagonal close-packed Se atoms where Pt atoms occupy the octahedral sites in alternative Se layers. The adjacent unoccupied Se layers are held together by weak van der Waals interactions, which emphasize the 2D nature. The top and side views of the crystal structure of PtSe$_2$ (PtTe$_2$) are shown in Fig.~\ref{Fig1}(b,c). The hexagonal bulk and projected surface Brillouin zones (BZs) are shown in Fig.~\ref{Fig1}(d) where high-symmetry points and Dirac points are also indicated.

Figure 2 shows the characterization of the PtSe$_2$ single crystals as compared to PtTe$_2$. Figure 2(a,b) shows the X-ray diffraction (XRD) patterns. Sharp (00l) diffraction peaks are observed.  The larger diffraction angles for corresponding (00l) peaks of PtSe$_2$ as compared to those of PtTe$_2$ suggests that PtSe$_2$ has a smaller layer separation. The extracted out-of-plane lattice constants are c = 5.07 $\pm$ 0.01 $\AA$ for PtSe$_2$ and c = 5.21 $\pm$ 0.01 $\AA$ for PtTe$_2$. The in-plane lattice constants are a = b = 3.727 $\AA$ for PtSe$_2$ and a = b = 4.0242 $\AA$ for PtTe$_2$ \cite{Kliche1985Far}. Figure \ref{Fig2}(c,d) shows the low energy electron diffraction (LEED) pattern of the PtSe$_2$ and PtTe$_2$ single crystals after cleaving in ultra-high vacuum chamber. Sharp diffraction spots with hexagonal symmetry are observed on the entire sample surface, indicating the high quality, homogeneous single crystals. Figure \ref{Fig2}(e,f) shows the temperature dependent resistivity of PtSe$_2$ and PtTe$_2$. A metallic behavior is observed in both samples from 380 K to 5 K with a high residual resistance ratio (RRR) of 184 for PtSe$_2$ and 29  for PtTe$_2$. The large interlayer spacing of PtSe$_2$ leads to changes in the lattice vibrations. The Raman spectrum in Fig.~\ref{Fig2}(g) shows the E$_g$ and A$_{1g}$ modes at $\sim$ 175 cm$^{-1}$ and 205 cm$^{-1}$ respectively for PtSe$_2$, which are at much larger energies compared to 110 cm$^{-1}$ and 157 cm$^{-1}$ in PtTe$_2$ (Fig.~\ref{Fig2}(h)).  The E$_g$  mode involves the in-plane vibration with the upper and the bottom Se atoms moving in opposite directions, and the A$_{1g}$ mode corresponds to the out-of-plane vibration of Se atoms.  The blue shift of the Raman modes in PtSe$_2$ indicates a stronger interaction than PtTe$_2$ \cite{ptse2Raman}, which is in good agreement with the smaller layer separation.

\begin{figure*}
  \centering
  \includegraphics[width=18cm]{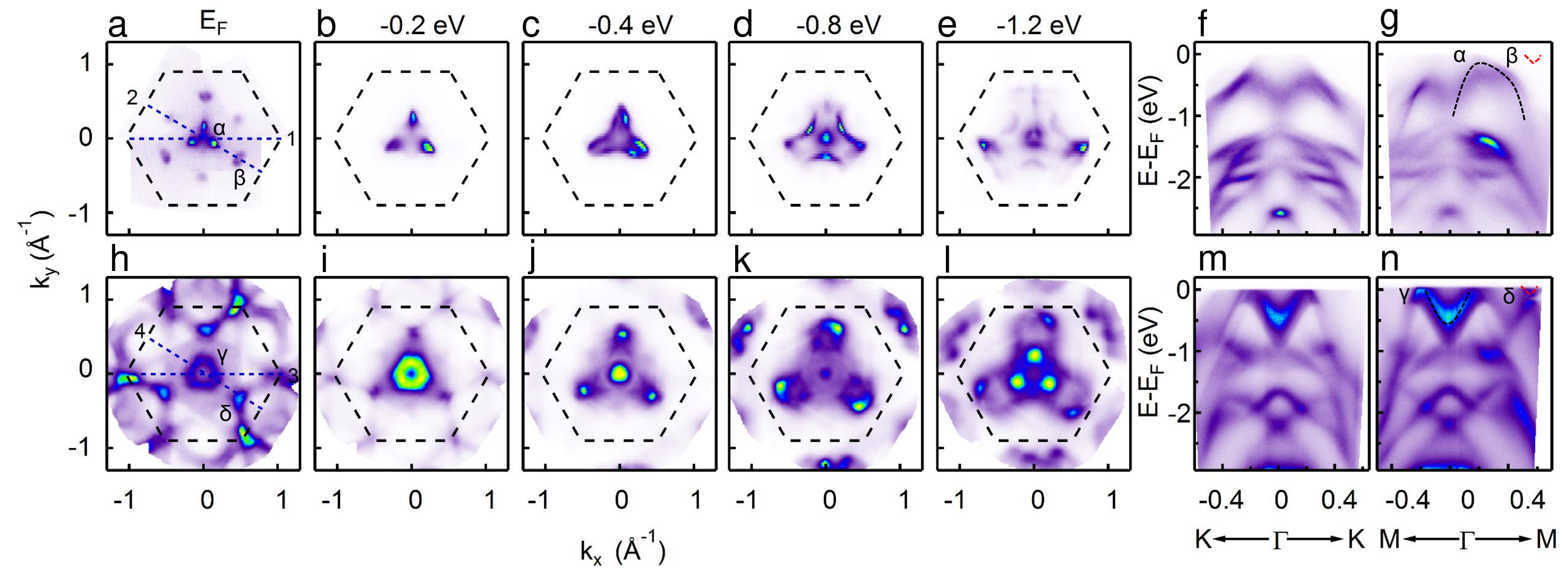}
  \caption{\textbf{Electronic structures of  PtSe$_2$ and PtTe$_2$.} (a-e) Intensity maps at constant energies from E$_F$ to -1.2 eV of PtSe$_2$ at a photon energy of 21 eV (k$_z$=0.29 c$^*$). Black dashed line indicates surface BZ, the locations of two momentum cuts are marked as blue dash lines 1, 2. (f,g) show the high-symmetry direction cuts 1 and 2 in (a), respectively. (h-l) Intensity maps at constant energies from E$_F$ to -1.2 eV of PtTe$_2$ at a photon energy of 21.2eV (k$_z$=0.3 c$^*$), the locations of two momentum cuts are marked as blue dash lines 3, 4. (m,n) show the high-symmetry direction cuts 3 and 4 in (h), respectively.
}\label{Fig4}
\end{figure*}

 The most important characteristic of a type-II Dirac semimetal is the tilted 3D Dirac cone in the band structure. To reveal the 3D Dirac cone, we measure the dispersions around k$_{\parallel}$ =0 at different k$_z$ values using different photon energies.  A conical dispersion is observed along the in-plane directions parallel to the $\Gamma$-K (Fig.~\ref{Fig3}(a)) and $\Gamma$-M (Fig.~\ref{Fig3}(b)) directions at k$_z$ = 0.37 c$^*$ (c$^*=2\pi/c$) of the reduced BZ.  Along the out-of-plane direction (Fig.~\ref{Fig3}(c)), the dispersion shows a highly tilted Dirac cone crossing at the same k$_z$ value. The band dispersions of PtSe$_2$ are overall similar to those of PtTe$_2$ (Fig.~\ref{Fig3}(e-h)) \cite{Yan2016Lorentz}, except that the Dirac points are at slightly different energies (-1.48 eV compared to -0.86 eV) and at different k$_z$ values (0.37 c$^*$ compared to 0.35 c$^*$). These differences are attributed to the different lattice constants.  In addition, compared to PtTe$_2$ where the intensity is suppressed at the crossing point showing a gap-like fetaure, PtSe$_2$ shows a 3D Dirac cone without any signature of gap opening at the Dirac point. The anisotropic 3D Dirac cones with the Dirac cone highly tilted along the k$_z$ direction provides direct evidence that it is a type-II Dirac semimetal.

The type-II Dirac fermions in PtSe$_2$ is also supported by first-principles calcualtions and symmetry analysis. The 3D Dirac cone corresponds to the band crossing along the in-plane S-D-T momentum path in calculated band structure in Fig.~\ref{Fig3}(d).  The Dirac cone is strongly tilted along the $\Gamma$-A direction, but not along the in-plane S-D-T path, which is the characteristic feature of the type-II Dirac fermions. The bulk Dirac cone of PtSe$_2$ is formed by two valence bands with Se-p orbitals labeled S$_1$ and S$_2$ in Fig.~\ref{Fig3}(d) (highlighted by red color), while that of PtTe$_2$ is formed by two valence bands of Te-p orbitals, labeled as T$_1$ and T$_2$ in Fig.~\ref{Fig3}(h). As each band is doubly degenerate, the isolated symmetry-protected band crossing is four-fold degenerate. Group theory analysis shows that these two bands belong to different irreducible representations. More importantly, the C$_3$ rotational symmetry about the c axis prohibits hybridization between them and protects the 3D Dirac fermions.  In addition to the 3D Dirac cone, there also exists conical Z$_2$ surface states which are induced by the band inversion. The deep topological surface states are revealed in the ARPES data in Fig.~\ref{Fig3}(a,b) (pointed by green arrows), and they connect the gapped bulk bands S$_3$ and S$_4$ with opposite parities in Fig.~\ref{Fig3}(d). Similar type-II Dirac cone and deep topological surfaces also exist in PtTe$_2$ (Fig.~\ref{Fig3}(e-g)).

Figure 4 shows a comparison of the electronic structure between PtSe$_2$ and PtTe$_2$ to reveal their difference.  Figure 4(a-e) and (h-l) show the evolution of the constant energy maps from E$_F$ to -1.2 eV. The Fermi surface map of PtSe$_2$ shows three small elliptical pockets around the $\Gamma$ point (denoted as $\alpha$ in Fig.~4(a)) and six pockets located at the mid-point between the $\Gamma$ and M points, which are from the bottom of the $\beta$ band (Fig.~4(g)). Below -0.4 eV, there is an additional pocket emerging at the $\Gamma$ point (Fig.~4(c)), and this pocket evolves into the 3D Dirac point at k$_z$ = 0.37 c$^*$ discussed above. The Fermi surface map of PtTe$_2$ features a small triangular pocket (denoted as $\gamma$) and six small circles (denoted as $\delta$ ) with alternating brightness located near the mid-point between $\Gamma$ and M.
The dispersions along the $\Gamma$-K and $\Gamma$-M directions are shown in Fig. \ref{Fig4}(f, g) for PtSe$_2$ and (m, n)for PtTe$_2$. The corresponding bands which contribute to $\alpha$, $\beta$ in (a), and $\gamma$, $\delta$ in (h) Fermi surface sheets are noted. Similar band structure can been found in PdTe$_2$ \cite{ZXJCPL,HanJin2016Experimental}. One major difference in the surface topology of PtSe$_2$ and PtTe$_2$ is that the band at the mid-point between $\Gamma$ and M shows a gapped Dirac cone in PtTe$_2$ , while there is only a small electron pocket in PtSe$_2$. These gapped Dirac cones are nontrivial surface states due to the extra band inversion for the top two bands at the $\Gamma$ point in PtTe$_2$ (Fig.~\ref{Fig3}(h)). Such band inversion does not occue in PtSe$_2$, leading to different surface states.

\section{IV. CONCLUSIONS}
 To summarize, by performing a systematic study of PtSe$_2$ single crystals, we provide the experimental evidence for type-II Dirac fermions in PtSe$_2$. The difference in the lattice constant, vibrational modes and electronic structure is also presented. Our work paves the way for studying a number of similar type-II Dirac materials in the PtSe$_2$ class of transition metal dichalcogenides. This family of TMDs provide a platform for investigating novel electronic structures and topologic phase transition like doping-driven Lifshitz transition \cite{BernevigNature,Xu2015Structured} in the near future.

\section{ACKNOWLEDGMENTS}
This work is supported by the National Natural Science Foundation of China (Grant No.~11427903 and~11334006), Ministry of Science and Technology of China (Grant No.~2015CB921001,~2016YFA0301001 and~2016YFA0301004).

\bibliographystyle{apsrev4-1}

\bibliography{reference}

\end{document}